\documentclass[12pt]{article}
\usepackage{psfrag,color}
\begin{document}
\hfill \begin{minipage}{70mm}
Proceedings of the Conference\\
``Path Integrals from peV to TeV''\\ 
Firenze, August 1998\\ 
to be published by World Scientific
\end{minipage}

\vspace{2mm}
\begin{center}
{\Large \bf The Quantum Dissipative Villain Model}

\vspace{1mm}
{G. Falci$^{(1)}$, G. Giaquinta$^{(1)}$ and U. Weiss$^{(2)}$}

\vspace{1mm}
{\small $^{(1)}$Istituto di Fisica, Universit\`a di Catania, 
         95125 Catania (I)\\ \& Istituto Nazionale 
        per la Fisica della Materia -- E-mail: gfalci@ing.unict.it\\
$^{(2)}$II.Institut f\"ur Theoretische Physik, Universit\"at
        Stuttgart, 70550 Stuttgart (D) }
\end{center}

\abstract{
 We introduce the Quantum Dissipative Villain (QDV) model as a prototype
 model to study tunneling in dissipative quantum mechanics. 
 Dissipation is provided
 by a coupled linear environment. In the QDV model, 
 the discrete character of a 
 tunneling degree of freedom coupled to an environment is
 explicit, leading to a rich dual structure.
 We derive general {\em exact} mappings of the QDV model on several dual 
 discrete  representations, including pairs of {\em self-dual} models, 
 for general linear
 environments and arbitrary temperatures. Self-duality allows to write
 exact equations for each correlation function of each representation. 
 Analogies with the theory of classical network 
 transformations are also presented. Finally we discuss the
 fundamental character of the QDV model. For instance, the standard 
 Caldeira-Leggett model, which describes, e.g., mesoscopic Josephson 
 junctions in a circuit and many other physical systems, is a special 
 QDV model.
 The self-dual structure of the QDV model  allows then the exact generalization
 of the Schmid approximate self-duality to general linear environments and 
 arbitrary temperatures.}

\section{Introduction}
Quantum dissipative systems can be described in terms of system plus environment
hamiltonians\cite{kn:Weiss-98} where a special quantum variable $\varphi$ 
interacts with an environment of harmonic oscillators. 
The reduced dynamics of  $\varphi$ obtained by integrating out
the reservoir's degrees of freedom is studied. 
An effective euclidean action is found
which, for state-independent dissipation, reads
\begin{equation}
\label{eq:general-action}
{\cal S}[\varphi] \;=\; 
        {1 \over 2} \int_0^{\beta} d \tau\, d \tau^{\prime}\,
        \varphi(\tau) \, {\cal A}(\tau-\tau^{\prime}) \,\varphi(\tau^{\prime})
        \;+\;  \int_0^{\beta} d \tau \,{\cal V}(\varphi) \quad ,
\end{equation}  
where  ${\cal V}(\varphi)$ is the potential.
The kernel, whose Fourier transform (FT) is\cite{kn:revs}  
${\cal A}(\omega)\!=\! m \omega^2\! - \!\frac{1}{2}\alpha(\omega)$, 
with $\omega \!=\! 2 \pi n/\beta$,  
contains the kinetic ``mass'' term and the damping term $\alpha(\omega)$
subsuming the spectral properties of the bath coupling~\cite{kn:Weiss-98}.

If ${\cal V}(\varphi)$
describes a  tunneling problem then
the variable $\varphi$ has an underlying discrete character even if
it is defined to be continuous. For instance, the low energy properties
of the double-well potential ${\cal V}(\varphi) = V (\varphi^2 -
a^2)^2$ are approximately accounted for by a two-state system. 
However, it is impossible to find the exact relation
for a generic environment.$\!$ We now introduce a model which
displays the discrete character of the continuous variable
$\varphi$ by writing
\vspace{-1mm}
\begin{equation}
\label{eq:QDVM-potential}
   \Lambda^{-1}\! \sum_{\tau}  {\cal V}(\varphi_{\tau}) 
        \,\, -\ln \Bigl[ \sum_{\{m_\tau \}} 
                     \exp \bigl\{
                        {1 / (2 \Lambda)} \sum_{\tau} 
                         V \left( \varphi_{\tau}
                 - 2 \pi m_{\tau} \right)^2 \bigr\}\Bigr]
        -  {\cal J}_{\tau} \, \varphi_{\tau} \;, \;\;
\end{equation}
where we added a source ${\cal J}_{\tau}$.
Here we are considering a version of 
eq.(\ref{eq:general-action}) on a lattice with $N$ sites and spacing 
$\beta/N = 1/\Lambda$. We name it Quantum Dissipative Villain (QDV) model
because eq.(\ref{eq:QDVM-potential}) is 
the Villain approximation\cite{kn:Savit-80} of the potential   
$\,{\cal V}(\varphi) \,=\, - V \cos \varphi - {\cal J} \varphi \,$ in the
Caldeira-Leggett (CL) model\cite{kn:Caldeira-Leggett}.

\section{Dual representations of the QDVM}
\noindent {\bf $m$-representation}: \hspace{2mm} 
We integrate $\varphi$ in the partition function of the QDV model,
which becomes
${\cal Z}^{QDV} [{\cal J}] = {\cal Z}^{V} [{\cal J}] \cdot
{\cal Z}^{(m)} [{\cal J}^{(m)}] \,$.
Here $ {\cal Z}^{V}$ is the partition function of the damped
harmonic oscillator (DHO), and 
\vspace{-1mm}
$$
\hskip-1mm
  {\cal Z}^{(m)} [{\cal J}^{(m)}]\; = \; \sum_{\{m_\tau \}}  
\mbox{\large e}^{ -
{1 \over 2}
                     \sum_{\tau,\tau ^{\prime}}
                   m_{\tau}  \, 
                   \left( {2 \pi V \over \Lambda} \right)^2 \, \left[
                        {\Lambda \over V} \delta_{\tau,\tau^{\prime}} -
                        \Lambda^2 {\cal G}_{\tau-\tau^{\prime}}^{V} 
                               \right] \,
                   m_{\tau ^{\prime}} + 
                 \sum_{\tau}
                 m_{\tau} {J^{(m)}_{\tau} / \Lambda} 
         } 
$$
is a 1-D surface roughening model  with interacting
heights\cite{kn:falci} $m_{\tau}$. The source 
is 
${\cal J}^{(m)}_{\omega} = 
2 \pi V \Lambda \, {\cal G}_{\omega}^{V} \, {\cal J}_{\omega}$. 
and ${\cal G}_{\tau}^{V}$ is the Green's function 
of the discretized DHO, given by  
$\;
{\cal G}_{\omega}^{V} = \Theta(\Lambda - |\omega |) \;
[m \Lambda \omega^2 - 
                   \Lambda \, \alpha(\omega)/2 + \Lambda \, V]^{-1}
\,$, for large enough $\Lambda$.\\[2mm]
{\bf $e$-representation}: \hspace{2mm} We introduce
$e_{\tau} := m_{\tau+1}-m_{\tau}$ and 
rewrite ${\cal Z}^{(m)}$ as
\vspace{-1mm}
\begin{eqnarray}
\label{eq:Z-e-representation}
{\cal Z}^{(e)} [{\cal J}^{(e)}] \,&=&\,
\sum_{\{e_\tau \}}  
\mbox{\large e}^{-\,
{1 \over 2} \; 
                     \sum_{\tau,\tau ^{\prime}} \,
                   e_{\tau}  \; \Delta_{\tau - \tau^{\prime}} \;  
e_{\tau ^{\prime}}
                   \;+ \; \sum_{\tau}  \; 
                         {\cal J}_{\tau}^{(e)} e_{\tau} } \;,
\\
&& \hskip-18mm 
\Delta_{\omega} \;=\;
    (2 \pi V \Lambda/ \omega)^2
        \left[  {1 / V}  -  \Lambda {\cal G}_{\omega}^{V} \right]
   \; ; \hskip6mm
  {\cal J}_{\omega}^{(e)} \;=\; 
  {- 2 \pi i V \Lambda / \omega} \;
  {\cal G}_{\omega}^{V} \; {\cal J}_{\omega} \;,\quad
\end{eqnarray}
\vspace{-4mm}

\noindent
obtaining a gas of interacting charges\cite{kn:falci}  
$e_{\tau} \in ]-\infty,\infty[$.  \\[2mm]
{\bf $n$-representation}: \hspace{2mm} Another charge
representation 
can be obtained starting from the 
QDV model, performing a Poisson transformations (which changes $m \to n$) 
and then integrating out $\varphi$. We obtain\cite{kn:falci}
${\cal Z}^{QDV} [{\cal J}] = {\cal Z}^{0} [{\cal J}] \cdot
{\cal Z}^{(n)} [{\cal J}^{(n)}] \,$ where ${\cal Z}^{0}$ describes a
Brownian particle and  ${\cal Z}^{(n)}$ has the same structure of
${\cal Z}^{(e)}$ eq.(\ref{eq:Z-e-representation}), being a gas of 
$n_{\tau} \in ]-\infty,\infty[$ charges with interaction and source
given by
\vspace{-1mm}
\begin{equation}
\label{eq:interaction-n-representation}
{\cal D}_{\omega} \;=\;
        {\Lambda / V}  +  \Lambda^2 {\cal G}_{\omega}^{0}
  \; ; \hskip6mm
   {\cal J}_{\omega}^{(n)} = 
  i \, \Lambda \,  
  {\cal G}_{\omega}^{0}  {\cal J}_{\omega}\;.
\end{equation}

\section{Exact Self-duality}
The ${\cal Z}^{(e)}$ and ${\cal Z}^{(n)}$ represent {\em the same model}, 
with modified interaction and sources. This means that the QDV model has an  
exact
{\em self-dual} structure\cite{kn:falci}. A simple reformulation of this 
self-dual mapping is 
obtained if we introduce the functions
$\, 
\zeta^0(\omega) = |\omega | / (2 \pi) \, \Lambda  {\cal G}_{\omega}^{0} 
\,$
and 
$\,
\zeta(\omega) = \zeta^0(\omega) 
+  |\omega| / (2 \pi V)
\,$.
Then we  rewrite 
$\Lambda^{-1} {\cal D}_{\omega} = 2 \pi / |\omega| \, \zeta(\omega) \,$ 
and
$\Lambda^{-1} \Delta_{\omega} = 2 \pi / |\omega| \, [\zeta(\omega)]^{-1}$.
The transformations of the interaction and of the source are finally given by
\begin{equation}
\label{eq:self-duality}
\zeta(\omega) \;\; \longrightarrow \; \; 1/\zeta(\omega) \;,
\qquad\mbox{and}\qquad
       {\cal J}_{\omega}^{(n)} 
        \;=\; - \omega/|\omega| \;
        \; \zeta(\omega)  \; {\cal J}_{\omega}^{(e)} \;.
\end{equation}

We can also write exact relations between correlation
functions of the representations of the QDV model. 
For instance, the FT of the
correlation function $\langle \varphi_{\tau} \varphi_0 \rangle $
of the QDV model
is related to 
$ \langle n \,  n \rangle_{\omega} =: 
|\omega|/(2 \pi \Lambda) \, {\cal C}_{\omega}[\zeta]$ by
\begin{equation}
|\omega|/(2 \pi \Lambda) \; \langle \varphi \varphi \rangle_{\omega}
\; = \;
\zeta^0(\omega) \; 
\Bigl\{ 1  \;-\; \zeta^0(\omega) \; 
{\cal C}_{\omega}[\zeta({\omega})]  \Bigr\}  \;. 
\end{equation}
Using self-duality, the relation between the $e$-$e$ and $n$-$n$
correlation functions becomes
an exact equation for
${\cal C}$
$$
\zeta(\omega) \; {\cal C}_{\omega} [\zeta(\omega)]
\;+\;
\zeta^{-1}(\omega) \;
{\cal C}_{\omega} [\zeta^{-1}(\omega)]
\;=\; 1 \;.
$$

\begin{figure}[t]
\label{fig:dualcircuit}
\centerline{\resizebox{130mm}{!}{\includegraphics{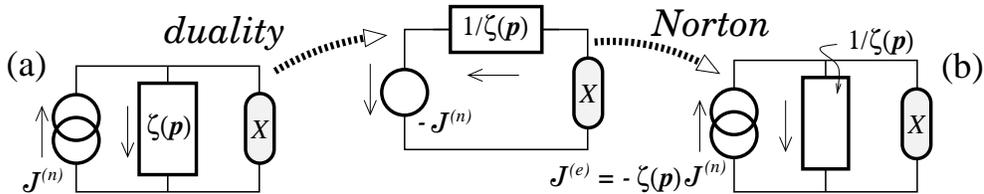}}}
\caption{
The $n$-representation corresponds to circuit (a), 
$\zeta(\mathbf{p})$ being the impedance seen by $\cal X$; 
the circuit (b) is obtained by a dual plus Norton
transformation and corresponds to the $e$-representation.
For the CL model we have $2 \pi m \leftrightarrow R_Q C$, 
$\alpha(\mathbf{p}) \leftrightarrow -
\mathbf{p} R_Q /[\pi Z(\mathbf{p})]$, $2 \pi V \leftrightarrow 
R_Q/L$, where $R_Q = h/(4e^2)$. 
Then $R_Q \, \zeta(\mathbf{p}) = [ Z^{-1}(\mathbf{p}) + \mathbf{p} C]^{-1} 
+ \mathbf{p} L$.}
\end{figure}

We have not yet  specified the environment, i.e. the function
$\zeta(\omega)$.
Both $\zeta(\omega)$ and $1/\zeta(\omega)$ have to be strictly
positive for $\omega \neq 0$ since otherwise  the integrations involved in the 
transformations cannot be performed.
Moreover the calculation of dynamic correlation functions
involves the analytic continuation $|\omega| \to \mathbf{p} \to i
\Omega + 0^+$, so it is desirable that $\zeta( \mathbf{p})$ is analytic
in ${\cal R}e \, \mathbf{p} > 0$. Thus we require that $\zeta( \mathbf{p})$
has the properties of the impedance of a linear
passive bipole\cite{kn:Chua-Desoer-Kuh}.
The analogy with network theory
involves also duality. Namely, eqs.(\ref{eq:self-duality})
for  ${\cal R}e \,  \mathbf{p} > 0$  can be reparaphrased
by associating to each charge representation a circuit with a non linear 
quantum component ${\cal X}$, the 
interaction $\zeta( \mathbf{p})$ corresponding  to the impedance seen by ${\cal X}$ 
and the current bias being ${\cal J}^{(.)}_{\mathbf{p}}$ (see fig.1). 
Then the quantum self-dual transformation for the charge models,
eqs.(\ref{eq:Z-e-representation},\ref{eq:interaction-n-representation},\ref{eq:self-duality}), correspond to 
transforming the linear elements and the source of the circuit 
using the known\cite{kn:Chua-Desoer-Kuh} classical dual and Norton 
transformations, while keeping 
unchanged the non-linear quantum component ${\cal X}$.

\section{Further developments}
The above results are significant in view of the fundamental character
of the QDV model. Indeed, the well known CL model can be obtained 
exactly from the QDV model in the continuum limit\cite{kn:falci} if 
$V \to \Lambda/[2 \ln(2\Lambda/V)]$. In other words this choice makes
{\em exact} the Villain approximation.  
In this case the low frequency limit of eq.(\ref{eq:self-duality}) 
reproduces both the approximate Schmid self-duality and the 
$\sigma \leftrightarrow - \sigma$ correspondence.\cite{kn:Schmid} The 
exact self-dual structure of the CL model was recently found for a 
special environment.\cite{kn:Fendley-Saleur} Here we find that it
holds true for arbitrary temperatures and general environments.  

When the CL model describes a mesoscopic Josephson junction\cite{kn:revs} 
in a circuit the analogy with network theory (see fig.1) 
becomes more stringent. 
The lowest order 
in the Coulomb-gas 
representation\cite{kn:falci} of
the $n$-model is the standard theory of the ``influence of the
environment''\cite{kn:revs2,kn:Weiss-98} and calculations can be performed
numerically for any external impedance. The same 
can be done 
for the lowest order in the 
$e$-representation, which corresponds to a single-instanton 
contribution\cite{kn:Weiss-98,kn:revs}.
Duality network relations for a purely resistive environment,
justified by the results of Schmid\cite{kn:Schmid} have been used 
for mesoscopic junctions 
since a long time\cite{kn:revs}. 
They are here substantiated and generalized.

\section*{Acknowledgments}
G.F. acknowledges R. Fazio and M. Annino
for discussions and suggestions, 
EU (TMR - FMRX CT 960042) and INFM (PRA-QTMD) for support.

\end{document}